\documentclass[aps]{revtex4-2}
\usepackage[pdftex]{graphicx}
\usepackage{amsmath}
\usepackage{amssymb}

\begin{document}

\title{Rescaling invariance and anomalous energy transport\\
	 in a small vertical column of grains}

	\author{A. Gnoli$^1$,  G. Pontuale$^2$, A. Puglisi$^1$ and A. Petri$^{1,3}$}\email[Corresponding author: ]{alberto.petri@isc.cnr.it}
\affiliation{$^1$CNR-Istituto Sistemi Complessi, Dipartimento di Fisica, Universit\`a Sapienza, P.le A. Moro, I-00185 Rome, Italy}
\affiliation{$^2$Council for Agricultural Research and Economics (CREA-FL), Via Valle della Quistione 27, I-00166 Rome, Italy}
\affiliation{$^3$ Enrico Fermi Research Center (CREF), via Panisperna 89A, 00184 Rome, Italy}
\begin{abstract}
It is well known that energy dissipation and finite size can deeply affect the dynamics of granular matter, often making usual hydrodynamic approaches problematic. Here we report on the experimental investigation of a small model system, made of ten beads constrained into a 1-d geometry by a narrow vertical pipe and shaken at the base by a piston excited by a periodic wave.     Recording the beads motion with high frame rate camera allows to investigate in detail the microscopic dynamics and test hydrodynamic and kinetic models. Varying the energy we explore different regimes from fully fluidized to the edge of condensation, observing   good hydrodynamic behavior down to the edge of fluidization, despite the small system size. Density and temperature fields for different system energies can be collapsed by suitable space and time rescaling, and the expected constitutive equation holds very well when the  particle diameter is considered. At the same time the balance between dissipated and fed energy is not well described by commonly adopted dependence, due to  the up-down symmetry breaking. Our observations, supported by the measured particle velocity distributions, show a different phenomenological temperature dependence,  which yields equation solutions in agreement with experimental results.    
\end{abstract}
\maketitle
\section{I. Introduction}
Granular matter can display a variety of behaviors  \cite{Jaeger1996,Puglisi2014}, from quasi-solid to fluid-like states, for which  effective descriptions exist in a limited number of regimes.  At the same time it is also a paradigmatic representation of a dissipative system  far from equilibrium and as such it is often  explored as a model system.  In the absence of external forces grain motion, even if initially present, eventually comes to an end because of the inelastic and frictional interparticle collisions. Otherwise it  can be sustained by continuous  energy supply.
Among many others, a  main issue  is whether and within which limits such situations can be described by hydrodynamics, where  that  granularity disappears and  the system state is defined by continuous  fields  in terms of local  averages of  quantities like velocity, granular temperature  and density.

Hydrodynamic descriptions of granular flow have been developed with some success (see e.g. the reviews in \cite{Goldhrisch2003,Campbell2006}). Being evident the advantages of such a description,   it can fail for several reasons. Among them, besides the discrete nature of the system components, there is the energy dissipation due to intergrain collisions and friction, that can generate strong gradients and  space-velocity correlations and  lead to clusterization \cite{McNamara1992,Li1995,Sela1995,Puglisi1998}.  To this respect it is critical the way in which  energy is fed \cite{Puglisi1998},  since it strongly influences dynamics through the way energy is redistributed \cite{Li1995}. But also
a homogeneous fluidized state can become unstable with respect to small density perturbations and  evolve so that a dilute granular fluid co-exists with much denser solid-like clusters \cite{Lu2022}. To address such situations it is often necessary to introduce  more  complex quantities, like variable viscosity, additive  diffusive terms etc. \cite{Campbell2006}, making applications problematic in several circumstances and
stimulating the formulation of computational methods based on effective interaction terms, like e.g. smoothed-particle hydrodynamics \cite{Szewc2016,Zhang2022,Xu2023}.

Granular hydrodynamics can represent a problem even in one dimension, as  shown in the seminal work by Li and Kadanoff \cite{Li1995} where a  system  can end in a static state because  grains clusterize far from the energy source.  One-dimensional systems are important also for the understanding of granular hydrodynamics in higher dimension, as stressed by Sela and Goldhrisch  \cite{Sela1995}.  In addition,  1--$d$ and quasi-1--$d$ granular systems  represent simplified situations to investigate  phenomena, like  e.g. wave transmission
\cite{Misra2019,Taghizadeh2021,Zhang2021,Jiao2023}. Their properties can  be of some relevance in the field of  active matter, where one dimensional systems are often considered \cite{Locatelli2015,Barberis2019,Dolay2020,Illien2020,Bar2020,Caprini2020,Banerjee2022}, and 
interesting  for applications, as for instance in granular dampers \cite{Ferreyra2021,Zhou2023}.

 Most of work on granular  hydrodynamics is based on calculations and numerical simulations, with sometimes disagreeing conclusions.  In 1--$d$   hydrodynamics  has been studied analytically under various conditions and different ways of  feeding  energy, generally for well fluidized, and  large systems \cite{Haff1983,Sela1995,Grossman1996,Grossman1997,Ramirez1999,Brey2001,Bromberg2003}, stimulating  a number of  simulations  \cite{McNamara1992,Bernu1994,Li1995,Sela1995,Grossman1996,Kadanoff1999,Goldhrisch1999,Luding1997,Puglisi1998,Soto1999,Alexeev2001,Baldassarri2001,Bromberg2003,Cecconi2004,carrillo2008,Barroso2009,Eshuis2009,Blackmore2014,Windows-Yule2017,Baldassarri2018}.
There are very few experimental works that, rather than verifying hydrodynamic behavior, are  generally  mainly focused  on collective dynamics  and  specific  phenomena,  like inversions or Leidenfrost effects,  among them \cite{Luding1997,Bocquet2001,Perez2008,Johnson2011,Lumay2013,Pontuale2016,Galvez2018}.

Far from giving  a  general description, our work aims at testing hydrodynamics experimentally in the simple case of small 1--$d$ systems in a statistically stationary state. 
As  remarked  above, similar systems have been subject of several studies from the theoretical and numerical point of view, but very few experimental instances can be found.
Despite the unavoidable  limits of the experimental conditions, this work aims at broadening the knowledge of the field  by providing an experimental demonstration that even a  small system of grains  is very  well described in terms of continuous hydrodynamic density and temperature fields with well defined properties.   

In the following, Sec. II is devoted to describe the experiment and its parameters, anticipating that the field profiles observed for  different system energies can all be overlapped by   suitable time and space rescaling.  Measured hydrodynamic fields are shown in Sec. III,   where  a state equation of the Van der Waals type  is successfully tested. 
In Sec. IV,  energy dissipation and current are experimentally measured, finding that the latter does not agree with what resulting from its  expression  usually adopted in terms of field.
On the base of experimental observations a different  expression for the temperature dependence is formulated,  which is also connected with the asymmetry induced by gravity in the velocity distribution of the grains.  In  Sec. V this expression is  employed in the hydrodynamic equations,  yielding  solutions in agreement with the observed fields. Final considerations are 
contained in Sec. VI, while  more  experimental and procedural details can be found in the Supplementary Information (SI).

\section{II. Experimental parameters and rescaling}
We have investigated a set of  $N$=10 identical steel beads of diameter $d$ = 4 mm,  restitution coefficient $\epsilon \simeq 0.92$, constrained to move in a vertical pipe 
 (Fig. S1 in SI). 
The energy is supplied to the system by a oscillating piston that hits  vertically the lowest grain,  and gravity prevents to reach  absorbing states with collapsed grains.  Changing the amount of fed energy allows  to explore different regimes.
The piston is driven sinusoidally  at a  frequency  $f=$  30 Hz, and  the grain motion is recorded by a video camera at 480 fps and digital images are processed to reconstruct the trajectories of the  center of mass of each bead (see SI). Being collisions substantially central, spin motion has not been taken into account. 
In the following, lengths and times will be converted from pixel and frames  into millimeters  and seconds, and the grain mass will be taken adimensional  and set equal to 1 (experimental and data processing details are available in the SI).

Here we  report on a series of 7 experiments  in which the piston  amplitude varies  from $A=2.90$ mm to $A = 1.15$ mm, with corresponding  driving temperatures $T_0=(2\pi f A)^2$ reported in Tab.~I. 
This choice of parameters  allows to explore the behavior of the system from  well fluidized to almost collapsed states.  
 \begin{table}[h]
	\centering
	\begin{tabular}{c|c|c|c|c|c|c|c|}
		\label{driving}
		Set S & 1  & 2  & 3 & 4 & 5 & 6 & 7  \\
		\hline
		$T_0$ (mm/s)$^2\cdot 10^4$  & 30.22 & 28.52  & 22.20 &  16.67 &   14.21  &   9.86  & 4.82 	\\
		$\lambda$ (mm) 
	&	70.8 &69.1&	 62.6 &57.0 &54.5 &50.1 &44.9\\
		$\tau$ (s) $\cdot 10^{-2}$ 
	&	8.85  &8.84 & 7.99 & 7.62 & 7.45 & 7.14 & 6.77
	\end{tabular}
	\caption{Values of temperatures characterizing the piston motion in the experiments considered here.}
\end{table}

In hydrodynamics,  generic (e.g. dimensional) considerations usually allow one to identify several length and time scales characterizing the system, such that different systems may  display the same dynamics  after suitable rescaling  of the fields. This expectation has been extended to granular systems 
but it is difficult to prove on a general foot, due to the huge number of possible granular regimes.
For specific systems like  that at hand, characteristic scales have been introduced in theoretical approaches  \cite{Bernu1994,Bromberg2003,Eshuis2009}.
\\
A first important attainment of the present work is to show that fields observed at different $T_0$ can be collapsed to a same profile if the finite size of the particles is taken into account in a suitable way. There are  different relevant scales in the system. In the present case,  being the supplied energy the varying parameter,  the relevant space and time scales  (remind that $m=1$) are related to the source temperature $T_0$, and without loss of generality can be taken respectively proportional to $\lambda=T_0/g$   and  $\tau=\sqrt{\lambda/g}$ \cite{Haff1983,Bromberg2003}.
However, experimental data show that in the present case this choice does not produce a good field collapse, which instead, is obtained rescaling by 
\begin{equation} 
\label{eq:scaling}
\lambda=T_0/g +Nd,
\end{equation}   which implicitly modifies also  $\tau$.  
Such dependence  appears natural since , taken two bead columns at rest made of different number $N$ with different diameter $d$ (identical in each column), they can be made look the same by measuring space in units of length  $Nd$. We test the validity of  this expression in  the following by considering the rescaled fields.  Notice that also  $\tau=\lambda/\sqrt{T_0}$ is a possible definition. The two choices for  $\tau$, respectively inertial and ballistic, coincide only for $d=0$ but  in the present case the second choice performs worse. 
\\

\section{III. Hydrodynamic description and fields}
Derivation of hydrodynamic equations for granular flow  has been performed in different ways, situations, and dimensions (see eg. \cite{Puglisi2014} and refs. therein). In their general form they are akin to those for true fluids but also account for the energy dissipated in collisions. Naming $z$ the only coordinate,
they have the form:
\begin{eqnarray}
\partial_t \rho & = & -\partial_z (\rho v) \label{euler1}\\
\rho\, \partial_t u & = & -\rho u \partial_z u -\rho g + \partial_z  P\\ \label{euler2}
\rho\, \partial_t T & = & -\rho  u \partial_z T -\partial_z J-P \partial_z u - W \label{euler3},
 \end{eqnarray}
where $u(z,t)$ and $\rho(z,t)$ are the velocity and density fields, $T(z,t)$ and $P(z,t)$ temperature and pressure, $g$ the gravity acceleration. 
These equations respectively  describe  mass conservation, momentum conservation (Euler equation) and energy balance. Here
$W$ is the rate of energy density dissipated in collisions and $J$  the energy current through the system.
 In the stationary state the first equation is trivially satisfied, and  the others dry to
 \begin{eqnarray}
 \partial_z  P-\rho g & = & 0 \label{stevino3}\\
 \partial_z J- W  & =  & 0 \label{sink}
  \end{eqnarray}
corresponding  respectively to the Stevino's law and to a continuity equation for the energy density. Suitable boundary conditions and constitutive relations are necessary to make closed the theory. Both ingredients are unknown in general. A main outcome of our study is the proposal of constitutive relations for pressure, energy current and dissipation rate.

\begin{figure}
	\begin{center}
		\includegraphics[width=0.4\textwidth]{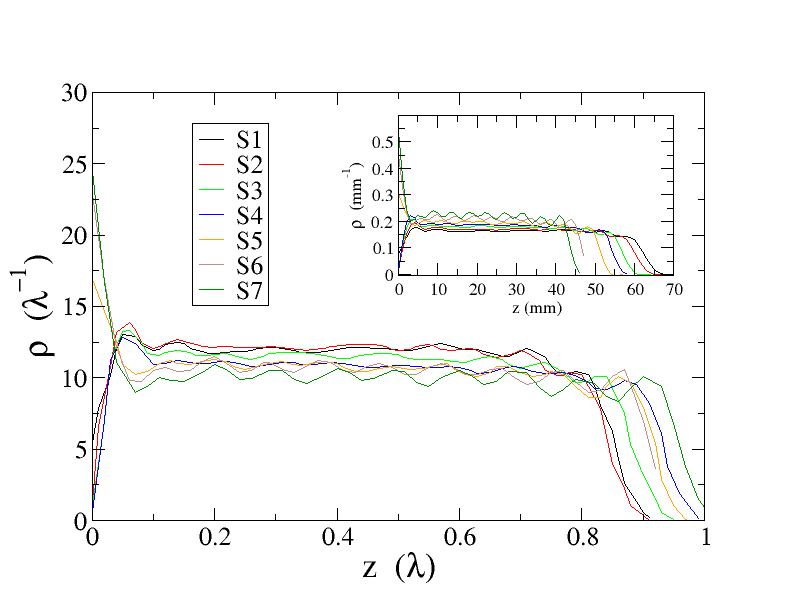}
		\includegraphics[width=0.4\textwidth]{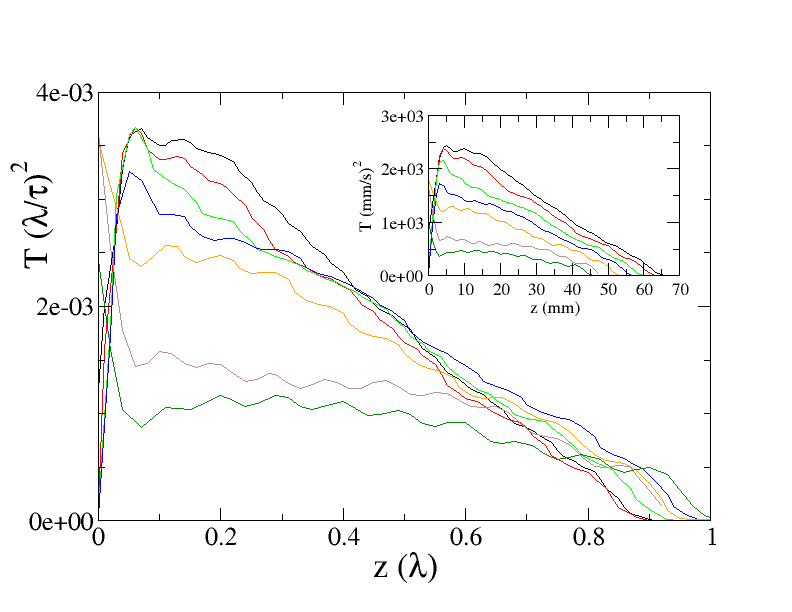}
		\caption{Density  (top) and temperature  (bottom) profiles rescaled by the characteristic scales $\lambda$ and $\tau$. Strong bead localization is visible at low energy.
		} 
		\label{fig:fields}
	\end{center}
\end{figure}

In order to verify  experimentally Eqs. (\ref{stevino3}) and (\ref{sink})  local stationary fields $\rho(z), v(z)$ and $T(z)$ have been evaluated from the  video recordings (see SI).
Figure~\ref{fig:fields} shows the  densities  $\rho$ (top) and  temperature $T$  (bottom) fields for the different sets considered. The main panels report the rescaled quantities:  $\rho \rightarrow \rho\lambda$, $T \rightarrow T \tau^2/\lambda^2$,  as functions of the rescaled height $z \rightarrow z/\lambda$. A good similarity is obtained for most of the cases: the fields appear rather smooth for sets of higher energy, while granularity  becomes  visible for decreasing  $T_0$, especially in sets S6 and S7. Density is very constant in the system bulk,  displaying  rather well equispaced relative maxima  with symmetric shape 
 in less fluidized systems, where  also increases close to the piston because of the low bead kinetic energy.     Rescaled temperatures decay about linearly far enough from the piston, displaying a common pattern  in regions  of increasing size for increasing  set energy,  sign that rescaling of hydrodynamics holds also for this granular systems if  particle diameter is suitably considered and system is fluid enough.  
Non rescaled densities and temperatures are plotted for comparison in the inset of the respective figures, showing real spatial extension and temperatures of the systems.

An important point concerns the system boundaries where  particular conditions act. On one side there is the energy source which affects the dynamics of the bottom  particle, and  consequently the fields in that region. On the other hand, the top particle   is free to jump  
and could deserve a separate ballistic dynamical description \cite{Haff1983}, but in the present case it does not show particular anomalies. It appears that the piston motion affects  some quantities. For instance, density and temperature vanish approaching z = 0  (Fig. \ref{fig:fields}), signaling a region of rarefaction and one can expect 
hydrodynamics to not hold in that region.  However, it must be noticed other quantities seem not affected, like pressure (Fig. \ref{fig:stevino}).
\\

\section{IV. Constitutive equations}
\begin{figure}
	\begin{center}
		\includegraphics[width=0.5\textwidth]{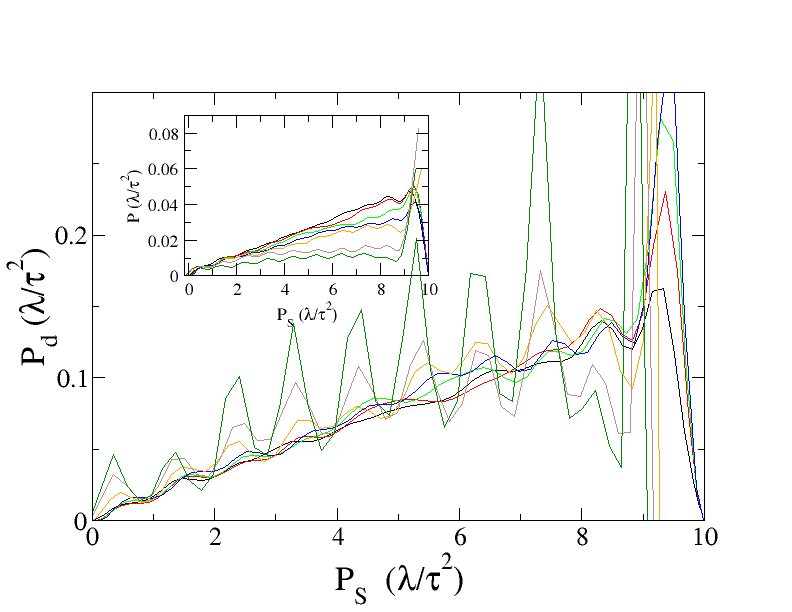}		
		\caption{Main panel: Test of Van der Waals-Tonks gas expression for pressure, $P_d$ (Eq. \ref{stevino3}), vs the  Stevino's law; Inset: the same using $P(z) = \rho(z) T(z)$ insted of $P_d$.}
		\label{fig:stevino}
	\end{center}
\end{figure}

From a variety of arguments (see e.~g.~\cite{Haff1983,Campbell1990,Sela1995}) one expects that in dilute situations the pressure $P$  follows: 
$P(z) = \rho(z) T(z)$,
namely it is proportional to the "internal" energy, like in a perfect gas. However,
 as anticipated, the finite diameter of the beads represents an important issue that must be considered, like in the  Van der Waals equation. An instance where it happens is represented by  the 1--$d$ Tonks gas of hard rods \cite{Tonks1936}, where pressure has the expression:
\begin{equation}
\label{Tonks}
P_d(z) =\frac{P(z)}{1-\frac{\rho(z)}{\rho_c}},
\end{equation}
with $\rho_c = \frac{N}{Nd}=\frac{1}{d}$. A similar  expression  has  been derived for a 1--$d$ model granular system \cite{Cecconi2004}.
The main panel of Fig.~\ref{fig:stevino} shows this quantity  vs
$P_S=g\int_z^\infty \rho(z) dz$, as suggested by   Eq. (\ref{stevino3}).
Both quantities are rescaled according to $\tau^2/\lambda$. 
It is seen that   pressure behaves smoothly and follows a linear trend in more energetic systems,  while in low energy systems display granularity, reflecting in large fluctuations which however do not change the average behavior. 
Notice that since the quantities on the two axes have the same dimension, the slope of the curves, which  turns out to be $\simeq 10^{-2}$,  do not depend on $T_0$  even without rescaling. 
 The importance of accounting for the finite diameter is demonstrated  in the inset of Fig. \ref{fig:stevino},  where   $P(z) = \rho(z) T(z)$ is plotted instead of $P_d$ resulting in a different slope for each set.
\\

Slightly different formulations can be found for the explicit expressions of $W$ and $J$ in Eq.~(\ref{sink}). From  kinematic arguments one expects  \cite{Li1995,Bromberg2003} $W= C_1 (1-\epsilon^2) \rho^2   T^{3/2}$, where $C_1$ is an adimensional constant. This expression neglects velocity-position correlations, which have been observed to invalidate it in some simulations  \cite{Mitarai2007}.  
Moreover, it has to suitably modify it to account for the finite particle diameter.  Following  the  derivation, it is easy to see that the modified expression reads:
\begin{equation}
\label{fsediss}
W =  C_1  (1-\epsilon^2)\dfrac{\rho^2 T^{3/2}}{1-\rho d},
\end{equation}
similarly to other cases.

\begin{figure}[h]
	\begin{center}
		\includegraphics[width=0.5\textwidth]{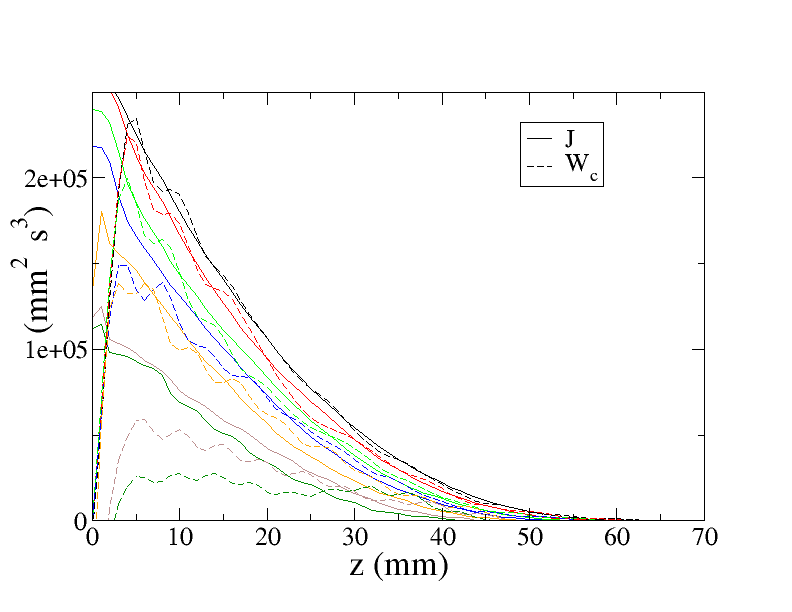}	
		\caption{Comparison of the  cumulated dissipated energy rate $W_c=\int W dz$ from Eq. (\ref{fsediss}) a with  the  energy  current $J$ (see text).}
		\label{fig:edissip}
	\end{center}
\end{figure}

We have tested  the expression (\ref{fsediss}) by considering  an explicit microscopic measure of the energy current \cite{Garzo1999},  which has its simple justification also in considering $\rho T$ as a kinetic charge and  multiplying it with its velocity $v$, as usual:
\begin{equation}
\label{ecurrent}
J(z)  = C_2 \rho(z) \langle  v^3(z) \rangle.
\end{equation}
 Evaluation of the constants $C_1$ and $C_2$ requires a detailed description of the kinetics and the related statistics, and is strongly dependent on a series of assumptions. Here we only adopt arbitrary values when necessary to compare different quantities. 
To avoid the noise  consequent to differentiation, we have  integrated  Eq. (\ref{sink}) with the boundary condition $J(\infty)=0$.  
 Moreover, being unknown whether  $J$ can  actually  be expressed in terms of fields, we have considered non rescaled quantities.
The results for all the  experimental sets are shown in Fig. \ref{fig:edissip}, where 
$W_c=\int_\infty^z W(z') dz'$.  
Apart from a  multiplicative constant the two quantities look to display rather close behavior far enough from the bottom. This is especially true for more energetic sets, where the curves are closer in a wider range,  down near the energy source, indicating that  the expression for $W$  in terms of fields taking into account the particle size, works  to a good extent.
The values of $C_2/C_1$ employed to make the two quantities comparably match range from $\approx 1.1$ for S1 to $\approx3.8$ for S7.

 Constitutive expressions relating the current $J$ with the fields have been obtained in various circumstances.
A natural expression, for small gradients, is 
  \begin{equation}
 J=\kappa \frac{\partial  T}{\partial z} + \mu \frac{\partial \rho}{\partial z}
 \label{nomcurrent}
 \end{equation}
 with $\kappa \simeq  \mu \propto T^{\frac{1}{2}}$.   We checked that this expression  does not work in the present case, as can be also easily seen by considering that, 
far enough from the piston, $T \simeq 1- const\cdot z$ and $\rho \simeq  const$ yield $J \approx (1- const \cdot z)^{\frac{1}{2}}$, well different from the curves in Fig. \ref{fig:edissip}. 

Expression (\ref{nomcurrent}) can be obtained by simple arguments  \cite{Haff1983}. More refined derivations based on  a Chapman-Enskog expansion  \cite{Brey1998,Garzo1999,Puglisi2014}, which  aims at expressing
 the local Probability Distribution  of Velocity (PDV) in terms of the other fields by expanding  small fluctuations, around a homogeneous  solution,  in terms of powers of the fields gradients. To find explicit expressions for the resulting  coefficients,  the  PDV is then usually expressed in terms of   polynomials that can   account  only for not too large perturbation of the Gauss-Maxwell  distribution. Moreover,  polynomials are  generally taken as functions of  $v^2$, assuming  a symmetrical PDV in force of the homogeneity and isotropy. The present one is of course not the case. Gravity and the way of supplying energy break  isotropy, enforcing a strong and permanent asymmetry of the PDV, as shown  in Fig.~\ref{fig:edissip} where in order to maintain the energy balance  the third moment of velocity looks persistent almost everywhere.

This is confirmed by Fig. \ref{fig:probvfS1}, where the local PDV  $p(z,v)$ evaluated from experimental data of set S1 is shown. Large asymmetries are seen. As can be expected from the behavior of $J$, they decrease towards the top of the column, but it can be verified that the velocity distributions are not Gaussian for any of the particle, even in almost symmetrical cases. This  feature is shared by all the experimental sets, including  low energy ones,  where asymmetry is weaker. It is also seen that close to the bottom the distribution is bimodal, a feature that could be spuriously due to the proximity of the piston, as already observed in different experiments \cite{Perez2008}.
\\

\section{V. Phenomenological description}
These last observations make problematic the description of the system   in terms of usual hydrodynamics fields $\rho$, $u$  and $T$. Starting  from a Gaussian PDV, asymmetries  can be accounted  in some cases by additional fields,  like for instance  in \cite{Sela1995}. However this is not always possible, like for instance in the presence of shear \cite{Campbell2006}. A different
approach  consists in considering the possibility that, despite  the "anormal" PDV,  current could be expressed in terms of the usual fields, although in not immediate way.  In this perspective it is of help to observe that as far as $J \simeq W_c$ then  $ J \approx T^{\frac{5}{2}}$.
In fact one can see that, at least far enough from the bottom,  from Fig.~\ref{fig:fields} one can assume $dT/dz \simeq const$, and hence $W_c = \int W dz = \int W  dT \frac{dz}{dT} \approx T^{\frac{5}{2}}$.

\begin{center}
\begin{figure}
		\includegraphics[width=15cm]{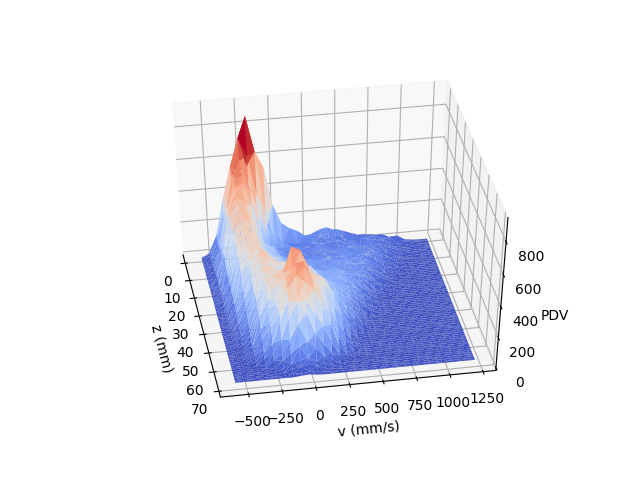}
		\caption{Probability distribution of the velocity field for the set S1.}
		\label{fig:probvfS1}
\end{figure}
\end{center}

 This looks at first sight weird, since from $J \propto \rho <v^3>$ and $\rho \simeq const$,  one would expect  $J \approx T^{\frac{3}{2}}$. This   implicitly assumes a linear dependence of the exponents characterizing  different moments: $<v^q> \propto T^{\frac{q}{2}}$, which  is not the case here. Notice that $J$ is close to $W_c$  in a wide range of  $z$ that increases down to $ z \approx 5$ mm in most energetic sets, implying a determinate, although non linear,  dependence between  the exponents of different moments of the PDV for a wide range of temperatures. This can be surprising since in Fig.~\ref{fig:probvfS1} PDV looks to change substantially in this range.
To test  whether $J \approx T^{5/2}$ is a reasonable guess, we use it in the hydrodynamic equations where, 
for such heuristic argument,  we neglect the excluded volume.  We also move to the Lagrangian frame where equations take a simpler form.  After rescaling by  $\lambda$   and $\tau$ (with $d=0$), transforming to the variable  $y=\int_0^z \rho(z')  dz'$ yields   \cite{Bromberg2003}:
\begin{eqnarray}
\rho T & = &(1-y) \label{laggas}\\
\rho \frac{\partial J}{\partial y} &  = &  W.
\label{lagsink}  \end{eqnarray}
Taking  $J \propto \rho T^{\frac{5}{2}}$ and $W \propto \rho^2 T^{\frac{3}{2}}$,  eliminating  $\rho$ through Eq.~(\ref{laggas}) yields:
\begin{equation}
\label{eq:alteq}	
\frac{\partial (1-y)T^{\frac{3}{2}} }{\partial y}=C(1-y) T^{\frac{1}{2}},
\end{equation}
 where $C$ is some  proportionality constant.
It is straightforward  to see that from the boundary condition 
$T(1)=0$  it  follows $T \propto (1-y)$ and  consequently, from Eq.~(\ref{laggas}), $\rho(y) \propto cost$. These
solutions qualitatively well agree with the experimental observations.   It can be worth mentioning that in \cite{Luding1997} some results from simulations with identical $N$ and $\epsilon$, and  $f=20$ Hz, also show  a linearly decaying temperature (see Fig. 5b) therein). Density is seen to decreases from bottom up (Fig. 5a)),  but with a trend to become more uniform for increasing $T_0$.

\begin{center}
	\begin{figure}
		\includegraphics[width=10cm]{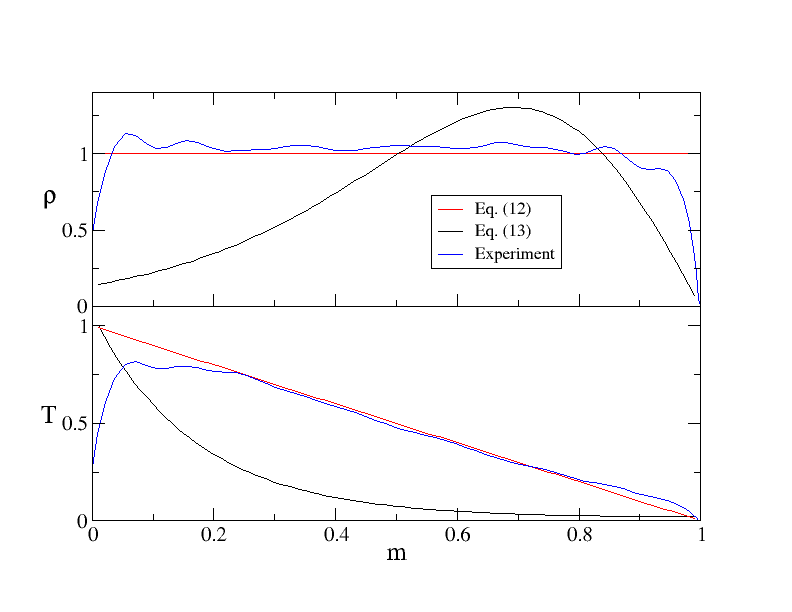}
		\caption{Experimentally observed density and temperature fields (set S1),  compared with the solutions of usual stationary  equation (\ref{eq:usualeq}),  and the alternative form
			(\ref{eq:alteq}) (fields are expressed in the Lagrangian coordinate and multiplied by arbitrary constants to ease the comparison). }
		\label{fig:comparison}
	\end{figure}
\end{center}

In order to compare our results with what expected from usual approaches \footnote{It can be worth remarking that slightly different equations can be derived, depending on the form of thermal conductivity} we have considered the equation \cite{Bromberg2003,Brey2001}:
\begin{equation}
\label{eq:usualeq}	
\frac{\partial (1-y)T^{\frac{1}{2}} }{\partial y}=\Lambda^2 (1-y) T^{\frac{1}{2}},
\end{equation}
whose solutions ,  with the boundary condition $dT/dy=0$ at $y=1 $  can be expressed  through the modified Bessel function of the first kind $I_0(x)$ as
\begin{equation*}
T(y)=\dfrac{I_0^2(\Lambda(1-y))}{I_0^2(\Lambda)}.\\	
\end{equation*}
Here $\Lambda=\frac{N\sqrt{\pi(1-\epsilon^2)}}{2} $ which in our case is $\simeq 3.47$. 
The resulting field profiles  are reported in Fig. \ref{fig:comparison}   together with the solutions of 
Eq.~(\ref{eq:alteq}) and 
those observed experimentally for set S1, confirming  that the usual field dependence of the energy current does not account for our observations, which are instead reproduced by the proposed form.

\section{VI. Discussion and conclusions}

The results here reported demonstrate that gravity is an efficient energy redistributor, but also a symmetry breaking factor which produces persistent asymmetry  of the velocity probability function in the stationary state, that we have measure experimentally. Such asymmetry can be  expected to persist also in higher dimension, and in larger systems, since it is essential to sustain the  upward energy flux,   and  it should  indeed increase because of the increasing dissipation.  On the other hand  it had already  been shown  
 that hydrodynamic description must include asymmetry even in the absence of symmetry breaking \cite{Sela1995}.

The  energy flux measured in the system is different from what expected by usual theories.  Adopting a   phenomenological expression derived from observations, we have found solutions of the hydrodynamic equations in agreement with the experimental results. 

A standard constitutive equation, that accounts for the finite bead diameter, has been observed to be well verified (in average even  in systems at the edge of condensation).  While this  had been  predicted by some granular theories, it  had  not yet been experimentally observed.  Finally, experiments show that many quantities can be rescaled by a  characteristic length  tied to $Nd$. This result and  the finite diameter  correction in the constitutive equation, as in other quantities,   are  expected to hold also for larger and higher dimensional  similar systems, i.e.  stacks of beads shaken from the bottom under gravity, provided that - in the directions perpendicular to the gravity - there are no inhomogeneities or instabilities such as convection etc.  On the contrary, the  expression adopted for the flux energy and the observed fields could be specific of the particular system investigated, and different in larger and higher dimensional systems. Nevertheless they bring about the lack of theory for systems like the one considered here, and  the necessity of considering odd moments of the velocity distribution.

\bibliographystyle{apsrev4-1}
\bibliography{bead.3}

\end{document}